\documentstyle{article}
\begin{document}

\def\ii{\'{\i}}
\def\bb{$\beta\beta_{2\nu}$}
\def\nd{$^{150}$Nd}
\def\sm{$^{150}$Sm}

\centerline{\bf DOUBLE BETA DECAY IN DEFORMED NUCLEI}

\smallskip

\noindent
\begin{center}
Jorge G. Hirsch$^1$, Victoria E. Cer\'on$^2$, Octavio Casta\~nos$^1$, Peter O. Hess$^1$\\
and Osvaldo Civitarese$^3$\\
{\small\it $^1$ Instituto de Ciencias Nucleares, UNAM, A.P. 70-543
Mexico 04510 D.F.}\\
{\small\it$^2$ Centro de Investigaciones Avanzadas en Ingenier\ii a Industrial,}\\
{\small\it UAEH, 42020 Pachuca, Mexico}\\
{\small\it$^3$ Departamento de F\'{\i}sica, UNLP, 
 c.c. 67 1900, La Plata, Argentina,}
\end{center}
\bigskip

\abstract{
The pseudo SU(3) approach has been used to describe many
low-lying rotational bands, as well as BE(2) and B(M1) intensities in rare
earth and actinide nuclei, both with even-even and odd-mass numbers 
\cite{Beu98,Rom98,Var00a,Beu00,Var00b,Pop00}.
The $\beta\beta$ half lives of some of these parent  nuclei to the ground
and excited states of the daughter ones were
evaluated for the two and zero neutrino emitting modes 
\cite{Cas94,Hir95a,Hir95b,Hir95c,Hir95d} using the pseudo SU(3) scheme.
The predictions were in good agreement with the available experimental data 
for $^{150}$Nd and $^{238}$U. 

The double electron capture half-lives of $^{156}$Dy, $^{162}$Er and $^{168}$Yb
were also studied using the same formalism \cite{Cer99}. 
The first nuclei was found to be the
best candidate for experimental detection, with a half-life of $\approx 10^{24}$ yr to
the first excited $0^+$ state in $^{156}$Gd.

There are strong selection rules  which restrict the two neutrino
mode of the $\beta\beta$ decay in some other nuclei, including  $^{160}$Gd.
Experimental limits for the $\beta\beta$ decay of $^{160}$Gd have been reported 
\cite{Bur95,Kob95}. 
Recently it was argued that the strong cancellation of the 2$\nu$ mode
in the $\beta\beta$ decay of $^{160}$Gd would suppress the background for the
detection of the 0$\nu$ mode \cite{Dan00}.

In the present contribution we extend the previous research 
\cite{Cas94,Hir95a,Hir95b,Hir95c,Hir95d}  evaluating
the $\beta\beta$ half lives of $^{160}$Gd using the pseudo SU(3) model. 
While the 2$\nu$ mode is forbidden
when the most probable occupations are considered, states with different
occupation numbers can be mixed through the pairing interaction. The amount
of this mixing is evaluated using perturbation theory. The possibility of
 observing
the $\beta\beta$ decay in $^{160}$Gd is discussed for both the 2$\nu$ and
0$\nu$ modes.}


\section{Introduction}

The calculation of two neutrino double beta decay matrix elements has proven to be
extremely sensitive to the details of the wave functions of the initial and final
nuclei \cite{Suh98}. While QRPA calculations are easy to perform, the uncertainties
in the residual particle-particle proton-neutron interaction strongly limit their
predictive power \cite{Suh98}. Shell model studies in the full {\em fp} shell
provide reliable matrix elements for $^{48}$Ca \cite{Rad96} and other light nuclei
\cite{Nak96}. Matrix elements for $^{76}$Ge, $^{82}$Se and $^{136}$Xe were obtained
using very large shell model spaces, which are however strongly truncated \cite{Cau96}.
For heavier nuclei standard shell model calculations are impracticable.

The pseudo SU(3) shell model \cite{Rat73,Dra84} is a microscopic model which allows the 
description of heavy deformed nuclei in the laboratory frame through the use
of a fermionic many-particle basis with good angular momentum.
The microscopic Hamiltonian employed includes single-particle energies as well as
pairing and quadrupole-quadrupole interactions. Most of the Hamiltonian parameters are fixed
by known systematics. Three ``rotor terms'' allow a fine tuning of the energy spectra and are
fitted for each nuclei. 
The pseudo SU(3) model has been used to describe many
low-lying rotational bands, as well as B(E2) and B(M1) intensities in rare
earth and actinide nuclei, both with even-even and odd-mass numbers 
\cite{Beu98,Rom98,Var00a,Beu00,Var00b,Pop00}. It was exhibited as a powerful  and 
predictive tool in the description of heavy deformed nuclei.

The $\beta\beta$ half lives of some of these parent  nuclei to the ground
and excited states of the daughter ones were
evaluated for the two and zero neutrino emitting modes 
\cite{Cas94,Hir95a,Hir95b,Hir95c,Hir95d} using the pseudo SU(3) scheme.
The predictions were in good agreement with the available experimental data 
for $^{150}$Nd and $^{238}$U. Predictions for double electron capture were
also performed. These results are briefly reviewed in the present contribution.

Extending the previous research 
\cite{Cas94,Hir95a,Hir95b,Hir95c,Hir95d}  
the $\beta\beta$ half lives of $^{160}$Gd are evaluated 
using the pseudo SU(3) model. 
While the 2$\nu$ mode is forbidden
when the most probable occupations are considered, states with different
occupation numbers can be mixed through the pairing interaction. The amount
of this mixing is evaluated using perturbation theory. 
The possibility of observing
the $\beta\beta$ decay in $^{160}$Gd is discussed for both the 2$\nu$ and
0$\nu$ modes.

\bigskip
\section{The pseudo SU(3) formalism}
\bigskip
\bigskip

In the pseudo $SU(3)$ shell-model coupling scheme~\cite{Rat73}, normal parity
orbitals $(\eta ,l,j)$ are identified with orbitals of a harmonic
oscillator of one
quanta less $\tilde \eta = \eta-1$. The set of orbitals with $\tilde j = j
= \tilde
l + \tilde s$,  pseudo spin $\tilde s =1/2$, and  pseudo orbital angular
momentum
$\tilde l$, define the so-called pseudo space. The orbitals with $j =
\tilde l \pm
1/2$ are nearly degenerate. For configurations of identical particles
occupying a
single $j$ orbital of abnormal parity, a convenient characterization of
states is
made by means of the seniority coupling scheme.

The many-particle states of $n_\alpha$ nucleons in a given shell
$\eta_\alpha$,
$\alpha = \nu $ or $\pi$, can be defined by the totally anti-symmetric
irreducible representations
$\{ 1^{n^N_\alpha}\} $ and $\{1^{n^A_\alpha}\}$ of unitary groups.
The dimensions of the normal $(N)$ parity space is
$\Omega^N_\alpha = (\tilde\eta_\alpha + 1) (\tilde\eta_\alpha +2)$ and
that of the unique $(A)$ space is $\Omega^A_\alpha =
2(\eta_\alpha +2)$ with the constraint
$n_\alpha = n^A_\alpha  + n^N_\alpha$.
Proton and neutron states are coupled to angular momentum $J^N$ and $J^A$
in both the normal and unique parity sectors, respectively. The
wave function of the many-particle state  with angular
momentum $J$ and projection $M$ is expressed as a direct product of the
normal and unique parity ones, as:

\begin{equation}
|J M > = \sum\limits_{J^N J^A} [|J^N> \otimes \: |J^A>]^J_M.
\end{equation}
Since we are interested in describing low-lying energy states, only pseudo spin
zero configurations are taken into account in the normal parity space and only
seniority zero configurations in the abnormal parity space.  This simplification
implies that $J^A_\pi = J^A_\nu = 0$. This is a strong assumption, but one that
is physically motivated and very useful for simplifying the calculations.

Double beta decay, when described in the pseudo SU(3) scheme, is
strongly dependent on the occupation numbers for protons and neutrons in
the normal and abnormal parity states:  $n^N_\pi, n^N_\nu, n^A_\pi,
n^A_\nu$~\cite{Cas94}.
These numbers are determined by filling the Nilsson levels from below, as
discussed in~\cite{Cas94}.

In the first series of papers \cite{Cas94,Hir95a,Hir95b,Hir95c,Hir95d} we evaluated 
the $\beta\beta$ matrix elements by taking into account only the leading SU(3) coupled 
proton-neutron irrep, which in recent calculation was shown to represent around
60 \% of the wave function in even-even Dy and Er isotopes \cite{Pop00}

\section {Allowed two neutrino $\beta^- \beta^-$ decays}

In all of the calculations only one active shell was allowed for protons, and
likewise, only one for neutrons. This is a very strong truncation. For the \bb
~decay this implies that only one uncorrelated Gamow-Teller transition is
allowed:
that which removes a neutron from a normal parity state with maximum angular
momentum and creates a proton in the intruder shell ($h^\nu_{9/2} \rightarrow
h^\pi_{11/2}$ in rare earth nuclei, $i^\nu_{11/2} \rightarrow i^\pi_{13/2}$ in
actinides). This unique Gamow-Teller transition controls the \bb ~decay.
Under these assumptions, if the occupation of the Nilsson levels is such that the number of
protons in the abnormal states does not change for the initial and final state
configurations, the decay is forbidden.

The published results for the six allowed two neutrino $ \beta^-\beta^-$ emitters are given 
in Table 1.
\begin{center}
\vskip -.5cm
$$
\begin{array}{ccc}
 \hbox{Transition}& \tau^{1/2}_{2\nu} [yr]  & \tau^{1/2}_{0\nu} [yr]\\
  \\
~~^{146}\hbox{Nd} \to ^{146}\hbox{Sm} ~~&~~ 2.1 \times 10^{31} &~~1.18 \times 10^{28}\\
~~ ^{148}\hbox{Nd} \to ^{148}\hbox{Sm} ~~&~~ 6.0 \times 10^{20} &~~6.75 \times 10^{24}\\
~~ ^{150}\hbox{Nd} \to ^{150}\hbox{Sm} ~~&~~6.0 \times 10^{18} &~~1.05 \times 10^{24} \\
~~^{186}\hbox{W} \to ^{186}\hbox{Os}~~ &~~ 6.1 \times 10^{24}  &~~5.13 \times 10^{25} \\
~~^{192}\hbox{Os} \to ^{192}\hbox{Pt} ~~& ~~9.0 \times 10^{25} &~~3.28 \times 10^{26}\\
~~^{238}\hbox{U} \to ^{238}\hbox{Pu} ~~& ~~1.4 \times 10^{21} &~~1.03\times 10^{24} \\ [12pt]
\end{array}
$$
\end{center}
\vskip -.5cm
Table 1: The calculated double
beta half-lives for the two-neutrino and the zero-neutrino

\bigskip

The agreement between the 
theoretical two neutrino half lives~\cite{Cas94} with the
available data for $^{150}$Nd ($\tau^{1/2}_{2\nu}= 9 (17)\times 10^{18}$ yr) 
and $^{238}$U ($\tau^{1/2}_{2\nu}=2 \times 10^{21}$ yr) is good.
For the theoretical $\beta\beta_{0\nu}$ half lives ~\cite{Hir95a} we assumed  $<m_\nu>~=$ 1eV.

\section {Allowed  double electron capture with two neutrino emission}

The $2\nu$ double electron capture (ECEC)
\begin{equation}
2e^{-}_b +(A, Z+2) \longrightarrow (A, Z) + 2\nu 
\end{equation}
has larger Q-values than the concurrent $\beta^+ \beta^+$ and $\beta^+$ EC processes, 
and has no Coulomb suppression but is very difficult to
detect, because only two X rays are emitted together with the neutrinos.

The double electron capture decay to excited states in the final nuclei
\begin{eqnarray}
(A,Z+2) + 2e^{-}_b \longrightarrow &(A, Z)^{*} + 2\nu
~~~~~~~~~~~~~~~  \\
& ~~~~^{\mid}\!\!\!\longrightarrow (A, Z) + 2 \gamma \nonumber
\end{eqnarray}
has been proposed as a good candidate to be measured \cite{Ver83,Bar94}.
The two gammas are far easier to detect than the X rays. 
A sensitivity close to $\sim 10^{22}$yr has been estimated for this type
of experiments \cite{Bar94}.

\begin{center}
\begin{tabular}{llc}
\multicolumn{2}{c}{Transition}  & $\tau^{1/2}$(yr) \\  
\\
$^{156}$Dy $\rightarrow ^{156}$Gd & $0^{+} \rightarrow 0^{+}(g.s.)$
& 2.74$\times 10^{22}$ \\
& $0^{+} \rightarrow 0^{+}(1)$ & 8.31$\times 10^{24}$ \\
& $0^{+} \rightarrow 0^{+}(2)$ & 1.08 $\times 10^{25}$ \\ 
\\
$^{162}$Er $\rightarrow ^{162}$Dy & $0^{+} \rightarrow 0^{+}(g.s.)$
& 2.85$\times10^{22}$\\
& $0^{+} \rightarrow 0^{+}(1)$ & 3.70$\times10^{27}$ \\ 
\\
$^{168}$Yb $\rightarrow ^{168}$Er & $0^{+} \rightarrow 0^{+}(g.s.)$ &
2.00$\times 10^{23}$ \\
& $0^{+} \rightarrow 0^{+}(1)$ & 5.36$\times 10^{33}$ \\ 
\end{tabular}
\end{center}
Table 2: Half-lives for the $ECEC_{2\nu}$ decay to the ground and excited
states of the final nuclei.\\

In Table 2 we present the $ECEC_{2\nu}$ decay of $^{156}$Dy, $^{162}$Er
and $^{168}$Yb to the ground and excited states of $^{156}$Gd, $^{162}$Dy
and $^{168}$Er respectively \cite{Cer99}. The kinematical factors $G_{2\nu}(\sigma)$
were evaluated following the prescriptions given in \cite{Doi88}. 
When the energy released in the decay to an excited stated ($2 E_\sigma$)
is small the available phase space $G_{2\nu}$ is strongly reduced, and
 the half life could be very large. It is the case in the
double electron capture decay to the first excited $0^+$ state in
$^{168}$Er. 
The nuclear matrix elements associated with the decay to the ground state
of the final nuclei  have values close to 0.05 -
0.06, a factor of 5 smaller than the assumption of Barabash \cite{Bar94},
and are similar for the three nuclei studied.  The nuclear matrix elements
to excited states  show a wide spread, being
close to those of the ground state for $^{156}$Dy, suppressed by a factor
5 for $^{162}$Er and by a factor 80 for $^{168}$Yb. While in general it is
confirmed that  deformed nuclei have smaller nuclear matrix elements than
spherical \cite{Cas94}, $^{156}$Dy appears to be the best candidate of
this group for experimental detection, with a half-life around $10^{24}$
years for the double electron capture to the first excited $0^+$ state.

\section{Forbidden two neutrino $\beta^- \beta^-$ decays}

As mentioned above, the number of nucleons in normal and unique parity orbitals is
determined by the filling of the deformed Nilsson orbitals. In this way the theory predicts 
the complete suppression of
the \bb ~decay for the following
five nuclei: $^{154}$Sm, $^{160}$Gd, $^{176}$Yb, $^{232}$Th and
$^{244}$Pu \cite{Cas94}.
It was expected that these forbidden decays would have, in
the best case, matrix elements that would be no greater than $20\%$ of
the allowed ones, resulting in at least one order of magnitude
reduction in the predicted half-life \cite{Hir98}.
Experimental limits for the $\beta\beta$ decay of $^{160}$Gd have been reported 
\cite{Bur95,Kob95}. 
Recently it was argued that the strong cancellation of the 2$\nu$ mode
in the $\beta\beta$ decay of $^{160}$Gd would suppress the background for the
detection of the 0$\nu$ mode \cite{Dan00}.

With a deformation of $\epsilon = 0.26$, the most probable occupations for the $^{160}$Gd 
14 valence protons are 8 in normal and 6 in unique parity orbitals, and for the
14 valence neutrons are 8 in normal and 6 in unique parity orbitals. The dominant component
of the wave function is 
\begin{eqnarray*}
|^{160}\hbox{Gd}, 0^+\rangle  = &
 | \ (h_{11/2})^6_\pi,  \ J^A_\pi =  0; \
(i_{13/2})^6_\nu ,\ J^A_\nu =  0 >_A \\
& | \{2^4\}_\pi (10,4)_\pi; \{2^4\}_\nu (18,4)_\nu; \ 1 (28,8) K=1, J = 0 >_N.
\end{eqnarray*}

Assuming a slightly larger deformation for $^{160}$Dy, the most probable occupations for 
16 valence protons are 10 in normal and 6 in unique parity orbitals, and for the
12 valence neutrons are 6 in normal and 6 in unique parity orbitals. The dominant component
of the wave function is 
\begin{eqnarray*}
|^{160}\hbox{Dy}, 0^+ (a)\rangle  = &
 | \ (h_{11/2})^6_\pi,  \ J^A_\pi =  0; \ (i_{13/2})^6_\nu ,\ J^A_\nu =  0 >_A \\
& | \{2^5\}_\pi (10,4)_\pi; \{2^3\}_\nu (18,0)_\nu; \ 1 (28,4) K=1, J = 0 >_N.
\end{eqnarray*}
The two neutrino double beta operator annihilates two neutrons and creates two protons with
{\em the same} spatial quantum numbers. It cannot connect the states $|^{160}\hbox{Gd},
 0^+\rangle$ and $|^{160}\hbox{Dy}, 0^+ (a)\rangle$ and the transition is forbidden.

However, the pairing interaction allows mixing between different occupations. In the deformed
single particle Nilsson scheme it takes about $\Delta E=$1.50 MeV to promote a pair of 
protons from the
last occupied normal parity orbital to the next intruder orbital. These excited state has
8 protons in normal and 8 in unique parity orbitals. Its wave function has the form
\begin{eqnarray*}
|^{160}\hbox{Dy}, 0^+ (b)\rangle  = &
 | \ (h_{11/2})^8_\pi,  \ J^A_\pi =  0; \ (i_{13/2})^6_\nu ,\ J^A_\nu =  0 >_A \\
& | \{2^4\}_\pi (10,4)_\pi; \{2^3\}_\nu (18,0)_\nu; \ 1 (28,4) K=1, J = 0 >_N.
\end{eqnarray*}
The two neutrino double beta decay is allowed to these excited state.

Using perturbation theory the $^{160}$Dy wave function is
$$|^{160}\hbox{Dy}, 0^+ \rangle =  a \,|^{160}\hbox{Dy}, 0^+ (a)\rangle  + 
b \,|^{160}\hbox{Dy}, 0^+ (b)\rangle ,$$
with $a^2 + b^2 = 1$ and 
$$ b = \langle ^{160}\hbox{Dy}, 0^+ (b) \,|\, H_{\hbox{pairing}} \,|\,^{160}\hbox{Dy}, 
0^+ (a)\rangle \, / \, \Delta E .$$

Using standard SU(3) techniques we obtain \cite{Hir01} a = 0.972, b = 0.233.
For the two neutrino mode the half-life is
$ \tau^{1/2}_{2\nu}$ $[ ^{156}$Dy $\rightarrow ^{156}$Gd, $0^{+} \rightarrow 0^{+}(g.s.)]$
= 1.65$\times 10^{22}$ yr. It is delayed by one order of magnitude (compared with other
nuclei with similar Q$_{\beta\beta}$ values) due to the presence of $b^2$ in the nuclear
transition matrix element.
The zero neutrino double beta transition is allowed to both states in $^{160}$Dy due to the
presence of the neutrino potential in the transition operator. Assuming again $<m_\nu>~=$ 1eV
the half-life is  $ \tau^{1/2}_{0\nu}$ $[ ^{156}$Dy $\rightarrow ^{156}$Gd,  
$0^{+} \rightarrow 0^{+}(g.s.)]$ = 3.96$\times 10^{25}$ yr \cite{Hir01}. 

The suppression of the 2$\nu$ mode would facilitate the search for the zero neutrino mode in
the double beta decay of $^{156}$Dy, as it was mentioned in \cite{Dan00}. However, the
value of the zero neutrino half-life is larger than the one expected in this reference.

\section{Conclusions}
The pseudo SU(3) shell model allows for a fully microscopic description of heavy
deformed nuclei in the laboratory frame, in a basis having good angular momentum and
particle number. It has proven to be a powerful and predictive technique
to describe the low energy spectra and electromagnetic transitions in rare earths and
actinide isotopes. It has also been applied to the description of the double beta decay in
these nuclei. 

The model Hamiltonian includes realistic single particle energies, 
quadrupole-quadrupole
and pairing interactions, whose strengths are fixed using known systematics. It also has
three rotor terms which a allow a fine tuning of the energy spectra. 
The wave functions mix different SU(3) irreps, while in even-even nuclei 
there is always one leading irrep which contributes with about 60\%
of the total wave function. 

Using only these leading irreps the double beta transition matrix elements were
evaluated. Both the two neutrino and zero neutrino mode were studied, involving
decays to the ground state and to excited states in the daughter nuclei. It was
found that, at first order, the two neutrino double beta decay is forbidden for many
candidates. The double electron capture was also studied for three nuclei,
concluding that the ECEC of $^{156}$Dy could be detected in future experiments.

Including different occupation numbers in normal and unique parity
orbitals for protons and neutrons through pairing mixing it was possible to
evaluate the {\em forbidden} half life of $^{160}$Gd, which is suppressed by a factor 10
(compared with other nuclei with similar $Q_{\beta\beta}$ values). It could offer a window
to detect the zero neutrino mode which, having more decay channels, is less suppressed by 
deformation.

\section{Acknowledgments}

This work was supported in part by CONACyT (M\'exico) and CONICET
(Argentina).


\begin{thebibliography}{9}
\bibitem{Beu98} T. Beuschel, J. P. Draayer, D. Rompf, J. G. Hirsch,
Phys. Rev. {\bf C 57} (1998) 1233.
\bibitem{Rom98} D. Rompf, T. Beuschel, J. P. Draayer, W. Scheid, J. G. Hirsch,
Phys. Rev. {\bf C 57} (1998) 1703.
\bibitem{Var00a} C. Vargas, J. G. Hirsch, T. Beuschel, J. P. Draayer,
Phys. Rev. {\bf C 61} (2000) 31301.
\bibitem{Beu00} T. Beuschel, J.G. Hirsch, and J.P. Draayer,
Phys. Rev. {\bf C 61} (2000) 54307. 
\bibitem{Var00b} C.E. Vargas, J.G. Hirsch and J.P. Draayer,
Nucl. Phys. {\bf A 673} (2000) 219-237.
\bibitem{Pop00} G. Popa, J. G. Hirsch and J. P. Draayer,
Phys. Rev. {\bf C 62} (2000) 064313.
\bibitem{Cas94} O. Casta\~nos, J.G. Hirsch and P.O. Hess, Rev. Mex.
Fis. {\bf 39} Supl. 2 (1993) 29;  O. Casta\~nos, J.G. Hirsch, O.
Civitarese and P.O. Hess, Nucl. Phys. {\bf A 571} (1994) 276.
\bibitem{Hir95a} J.G. Hirsch, O. Casta\~nos and P.O. Hess, Nucl. Phys.
{\bf A 582} (1995) 124.
\bibitem{Hir95b} J.G. Hirsch, O. Casta\~nos, P.O. Hess and O.
Civitarese, Nucl. Phys. {\bf A 589} (1995) 445.
\bibitem{Hir95c} J.G. Hirsch, O. Casta\~nos, P.O. Hess and O. Civitarese,
 Phys. Rev. {\bf C 51} (1995) 2252.
\bibitem{Hir95d} J. G.  Hirsch, Rev. Mex. Fis. {\bf 41} Supl. 1 (1995) 81-89.
\bibitem{Cer99} V. E. Cer\'on and J.G. Hirsch, Phys. Lett. {\bf B 471} (1999) 1.
\bibitem{Bur95} S.F. Burachas, F.A. Danevich, Yu.G. Zdesenko, V.V. Kobychev, V.D. 
Ryzhikov, and V.I. Tretyak, Phys. At. Nucl.{\bf 58} (1995) 153.
\bibitem{Kob95} Masaaki Kobayashi, Shigeharu Kobayashi, Nucl. Phys. {\bf A 586} (1995) 
457. 
\bibitem{Dan00} F.A. Danevich, V.V. Kobychev, O.A. Ponkratenko, V.I. Tretyak, and Yu.G. 
Zdesenko, arXiv:nucl-ex/0011020.
\bibitem{Suh98} J. Suhonen, O. Civitarese, Phys. Rep. {\bf 300} (1998) 123.
\bibitem{Rad96} P.B. Radha et al., Phys. Rev. Lett. {\bf 76} (1996] 2642.
\bibitem{Nak96} H. Nakada, T. Sebe, and K. Muto, Nucl. Phys. {\bf A 607} (1996) 235.
\bibitem{Cau96} E. Caurier, F. Nowacki, A. Poves, and J. Retamosa, Phys. Rev. Lett.
{\bf 77} (1996) 1954.
\bibitem{Rat73}  R.D. Ratna Raju, J.P. Draayer and K.T. Hecht,
Nucl. Phys. {\bf A 202} (1973) 433; K.T. Hecht and A. Adler,  Nucl. Phys.
{\bf A 137}(1969) 129; A. Arima, M. Harvey and K. Shimizu, Phys. Lett.
{\bf B 30} (1969) 517.
\bibitem{Dra84} J.P. Draayer and K.J. Weeks,  Ann. Phys. {\bf 156} (1984)
41;   O. Casta\~nos, J.P. Draayer and Y. Leschber,  Ann. of
Phys. {\bf 180} (1987) 290.
\bibitem{Ver83} J.D. Vergados, Nucl. Phys. {\bf B 218} (1983) 109;
M. Doi, T. Kotani, E. Takasugi, Progr. Theo. Phys. Suppl.{\bf 83} (1985) 1.
\bibitem{Bar94} A. S. Barabash, JETP Lett. {\bf 59} (1994) 677.
\bibitem{Doi88} M. Doi, and T. Kotani, Phy. Rev. {\bf C 37} (1988) 2104;
M. Doi, and T. Kotani, Progr. Theo. Phys. {\bf 87} (1992) 1207.
\bibitem{Hir98} J.G. Hirsch et al, Czec. Journ. Phys. {\bf 48} (1998) 183.
\bibitem{Hir01} J.G. Hirsch, O. Casta\~nos, P.O. Hess and O. Civitarese, in preparation.
\end{thebibliography}
\end{document}